\documentclass[twocolumn]{article}
\usepackage{abstract}
\usepackage[margin=71pt]{geometry}
\usepackage{cite}
\usepackage{graphicx}
\usepackage{color}

\let\oldbibliography\thebibliography
\renewcommand{\thebibliography}[1]{%
  \oldbibliography{#1}%
  \setlength{\itemsep}{0pt}%
}

\begin{document}

\title{Periodic compression of an adiabatic gas:\\
Intermittency enhanced Fermi acceleration}

\author{Carl P. Dettmann\thanks{School of Mathematics, University of Bristol, Bristol BS8 1TW, United Kingdom} and Edson D. Leonel\thanks{Departamento de F\'isica - UNESP - Univ Estadual Paulista, Av.24A, 1515 - Bela Vista - CEP: 13506-900 - Rio Claro - SP - Brazil}}

\date{\today}

\twocolumn[
\begin{@twocolumnfalse}
\maketitle

\begin{abstract}
A gas of noninteracting particles diffuses in a lattice of pulsating scatterers.
In the finite horizon case with bounded distance between collisions and strongly
chaotic dynamics, the velocity growth (Fermi acceleration) is well described by a
master equation, leading to an asymptotic universal non-Maxwellian velocity
distribution scaling as $v\sim t$. The infinite horizon case has intermittent dynamics
which enhances the acceleration, leading to $v\sim t\ln t$ and a non-universal distribution.
\end{abstract}
\end{@twocolumnfalse}
]
\saythanks

Periodically forced thermally isolated systems exhibit many interesting phenomena
from stabilization~\cite{GMRF} to exponential acceleration~\cite{GRST}.  They are
of intense interest for trapped atom~\cite{SKHGMW}
and ion~\cite{OLABLW} experiments, as well as astrophysical problems such as transport
of comets~\cite{S11}.  Using a spatial coordinate as the independent variable,
they also describe transport in periodic structures~\cite{LCD}.  Typically there is unbounded
growth of the energy, the phenomenon of Fermi acceleration~\cite{F} (FA).  Very
recently, many researchers have sought analytical descriptions of energy
distributions in such systems~\cite{BDKP,KDPS,KDC,DK}.  In rather general circumstances a
Fokker-Planck (FP) equation can be derived, incorporating the average and variance of
the work per period~\cite{BDKP}.  The first example in~\cite{BDKP}, extensively investigated
elsewhere~\cite{DK,JS,KDC,KDPS,KPDCS,LRA99,LRA00}
consists of a particle, or equivalently gas of noninteracting particles,
moving freely in a container (``finite billiard'') or amongst obstacles
(``Lorentz gas'') with oscillating boundary but fixed volume.  Our main aim is to investigate
forced systems with oscillating volume, developing methods (applicable to general classes of
FA systems) to characterise and tame the resulting wild oscillations in the energy distribution.
We exhibit contrasting features of chaotic and intermittent regimes, including the paradoxical
effect that in the intermittent case, fewer collisions lead to greater acceleration.

Periodically oscillating billiard(-like) models exhibiting FA include the
1D bouncer~\cite{DL} and stochastic simplified Fermi-Ulam~\cite{KDC} models.
In the latter (and often elsewhere), the simplifying assumption of the static wall
approximation (SWA) was used, where the boundaries are fixed (hence trivially having
fixed volume) but the particle changes its velocity as if they were moving.
Many oscillating two-dimensional billiards have also been considered and lead to FA.
It is conjectured that this includes all chaotic geometries~\cite{LRA99,LRA00},
as well as the ellipse~\cite{LDS}.  The breathing case (fixed shape) has been studied in
detail~\cite{BR}, leading to slower growth of velocity than other typical models.
FA is normally prevented by dissipation in the dynamics, although
there are scaling laws relating the final energy to the strength of the dissipation~\cite{OVL}.

Jarzynski and Swiatecki~\cite{JS}, showed using moments that for fixed volume
time-dependent billiards, the eventual distribution of velocities is exponential, in contrast
to the Gaussian distribution of an equilibrium gas; this was confirmed numerically
in Ref.~\cite{BBS93}.  Jarzynski~\cite{J} then described an
FP equation approach for a slowly varying billiard (or fast particle) giving an explicit calculation
of the rates of increase of the energy and its variance;
this was later applied, with some
further approximation and  difficulties due to dynamical correlations, to a system with
oscillating volume~\cite{BJS96}. Bouchet, Cecconi and
Vulpiani~\cite{BCV} in an astrophysical context applied a linear
Boltzmann equation to obtain an exponential velocity distribution.  More recent
innovations have included a hopping wall approximation replacing the SWA~\cite{KPDCS},
and a Chapman-Kolmogorov equation replacing an FP equation~\cite{KDC}.  Here
we retain the simpler FP approach, but treat the wall collisions exactly.  Many of these
techniques are also relevant to stochastically moving
boundaries~\cite{DK,GMPS}.

\begin{figure}
\vspace{-45pt}
\centerline{\includegraphics[angle=270,width=230pt]{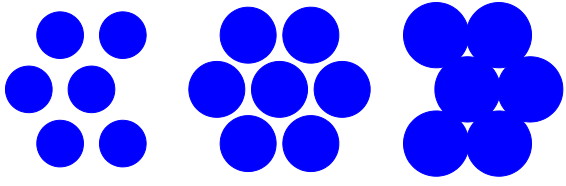}}
\vspace{-45pt}
\caption{\label{f:lor} The triangular Lorentz gas consists of a particle moving freely except
for reflections with a periodic array of obstacles.  Transport consists of three regimes: Infinite
horizon (left) with unbounded free path length, Finite horizon (center) with bounded
free paths but overall diffusion, and confined (right).}
\end{figure}

Our model is a two dimensional Lorentz gas,
a collection of circular scatterers in an extended domain.  Fixed
random~\cite{KPV} and periodic~\cite{D} scatterer
arrangements have been widely studied for the last century.  In the periodic case, the
transport regimes are infinite horizon (IH), finite horizon (FH) and confined (C) as
illustrated in Fig.~\ref{f:lor}.  In the IH and FH cases a particle can diffuse to arbitrarily
great distances; Green-Kubo formulae~\cite{EM08} express the diffusion coefficient as the
infinite time integral of the velocity autocorrelation function
$\langle v(t)v(0)\rangle$.  In the FH case this is believed to decay as $\exp(-Ct)$, while
in the IH case as $C/t$.  Thus for IH the integral diverges, leading to logarithmic
superdiffusion~\cite{D}.
In both cases the collision dynamics is strongly chaotic, and the
anomalous IH diffusion is due to long flights.

We place the scatterers on a triangular lattice, with each unit cell having unit area,
so the distance between the centres of neighbouring scatterers is
$(4/3)^{1/4}$.  The triangular Lorentz gas is IH for
$r<r_H=(3/64)^{1/4}\approx 0.465$, FH for  $r_H<r<r_I=(1/12)^{1/4}\approx 0.537$,
at which the scatterers start to intersect, and confined (C) for
$r_I<r<r_B=(4/27)^{1/4}\approx 0.620$ at which point the dynamics is
blocked as there is no space outside the scatterers.  The area available
to a billiard particle ${\cal A}(r)$ is
\begin{equation}\label{e:area}
\left\{\begin{array}{cc}
1-\pi r^2&r\leq r_I\\
1-r^2\left(\pi-6\arccos\frac{r_I}{r}+6\sqrt{\frac{r_I^2}{r^2}-\frac{r_I^4}{r^4}}\right)& r\geq r_I
\end{array}\right.
\end{equation}
 
Here we consider time-dependent scatterers, with radius $r(t)=R+A\sin t$
and boundary velocity $u(t)=r'(t)=A\cos t$.
There are several scenarios depending on $R_\pm=R\pm A$. Our
\{I,F,C\} notation indicates what regimes exist as time passes, so IFC indicates
that between infinite and confined times there is a finite horizon.

\begin{description}
\item[I] Infinite (horizon)  \hfill $R_+<r_H$
\item[IF] Infinite, finite \hfill $R_-<r_H<R_+<r_I$
\item[IFC] Infinite, finite, confined \hfill $R_-<r_H<r_I<R_+$
\item[F] Finite  \hfill $r_H<R_-<R_+<r_I$ 
\item[FC] Finite, confined \hfill $r_H<R_-<r_I<R_+$
\item[C] Confined \hfill $r_I<R_-$
\end{description}

Ref.~\cite{KDPS} has some discussion of a IF model (with fixed volume), denoting it
``dynamically infinite horizon.'' For the numerical simulations we choose $A=0.03$,
which allows all the above cases except IFC.  A Lorentz gas on a square
lattice has no finite horizon, and so exhibits regimes I, IC and C.

We first discuss FA for the finite or confined geometries.
The billiard particles move
freely, colliding with the scatterers according to~\cite{BR}
\begin{equation}\label{e:vvec}
{\bf v}_+={\bf v}_-+2{\bf n}(u-{\bf n}\cdot{\bf v}_-)
\end{equation}
where ${\bf v}_+$ (${\bf v}_-$) is the velocity immediately after (before)
the collision, $\bf n$ is an outward unit normal at the point of collision,
and we use $v_\pm=|\bf v_\pm|$. 
The incoming angle $\theta$ with respect
to the normal satisfies
$-{\bf n}\cdot{\bf v}_-=v_-\cos\theta$.
If a particle with $v_-<u$ is overtaken by the scatterer then $\theta>\pi/2$.  We define
$\theta\geq 0$ so there is a 1:1 relation between $\theta$ and the
outgoing speed $v_+$, thus each $\theta>0$ corresponds to two incoming directions.
Eq.~(\ref{e:vvec}) gives
\begin{equation}\label{e:v+}
v_+^2={\bf v_+\cdot v_+}=v_-^2+4uv_-\cos\theta+4u^2
\end{equation}
Thus the change in speed is the same sign as $u$ and of magnitude up to $2|u|$.

This system exhibits FA, and almost all initial
conditions to lead to unbounded speed; after sufficient time
$v$ exceeds all velocity scales set by the problem, including $|u|$ and the
lattice spacing times the oscillation frequency.
Thus the particles are effectively in a Lorentz gas with slowly varying radius, and as in
the static case, having exponential decay of time correlation functions.
The only quantity not randomised by the dynamics at short times is $v$,
a constant of motion for the static case.

Thus we may describe the system by a spatially homogeneous
distribution function
$f(v,t)\delta v$ giving the probability of observing a particle with speed
in the interval $[v,v+\delta v]$ at time $t$, hence normalised so
\begin{equation}\label{e:norm}
\int_0^\infty f(v,t) dv=1
\end{equation}
 for all $t$.  The probability of finding the particle in a
region of the full phase space is, under this assumption,
$f(v,t)\delta v\frac{\delta\psi}{2\pi}\frac{\delta x\delta y}{\cal A}$
where $\psi$ denotes the direction of the velocity, including the relevant
normalisation factors.  Here, and often later,
the time dependence of $r$ (and hence $\cal A$) has been suppressed for simplicity.

The distribution $f(v,t)$ evolves due to collisions with the scatterers,
which make small changes of order $u$ to the speed.  The collisions
depend on one distribution function and the known position and
velocity of the scatterers, so the treatment here is a continuous state
master equation, similar to the linear Boltzmann equations of Refs.~\cite{BCV,DK}
(but spatially homogeneous).
Correlations between collisions are neglected
(but can be included using results of Ref.~\cite{J}; see the appendix), but due
to the mixing (hence also ergodicity) property of the dynamics, a long sequence of
collisions has the same effect as a Markov chain with the correct probability distribution.

The general form of a master equation is
\begin{equation}
f_t(v,t)=\int [p(v,v')f(v',t)-p(v',v)f(v,t)] dv'
\end{equation}
where subscript $t$ (later $v$) is the partial derivative and
$p(v,v',t)$ is the collision rate for a collision taking $v'$ to $v$;
it has explicit time-dependence from the moving boundaries which is again
suppressed.  We need to find the
probability of a collision taking $v_-\in[v',v'+\delta v']$ to $v_+\in[v,v+\delta v]$
at a time in $[t,t+\delta t]$ by integrating the distribution over the set of
trajectories with the appropriate collision.

\begin{figure}
\vspace{-80pt}
\centerline{\includegraphics[angle=270,width=230pt]{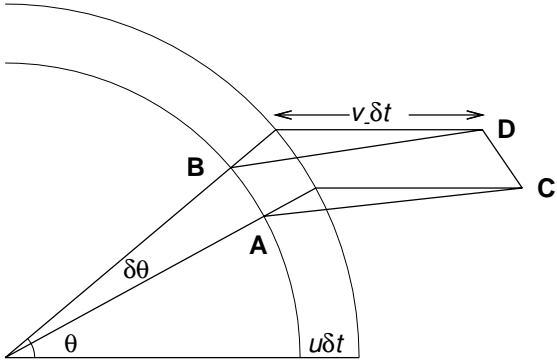}}
\vspace{0pt}
\caption{\label{f:diag} A particle moves in the negative $x$ direction.  The (almost)
parallelogram ABDC denotes the initial positions that will collide in time $[0,\delta t]$,
at location $[\theta,\theta+\delta\theta]$ taking the moving boundary into account.}
\end{figure}

For $r<r_I$ the cross section is independent of $\psi$, so we take $\psi=\pi$; larger radii
or different chaotic maps would need to take the $\psi$-dependence into account.
The trajectory hits the scatterer at time $t$
at a point $A(r(t)\cos\theta,r(t)\sin\theta)$; see Fig.~\ref{f:diag}.
A trajectory hitting the scatterer at angle $\theta+\delta \theta$
reaches it at $B(r(t)\cos(\theta+\delta\theta),r(t)\sin(\theta+\delta\theta)$.
To reach the scatterer at $t+\delta t$, the particle at time $t$
will be at $C(r(t+\delta t)\cos\theta+v_-\delta t,r(t+\delta t)\sin\theta)$
or $D(r(t+\delta t)\cos(\theta+\delta\theta)+v_-\delta t,r(t+\delta t)\sin(\theta+\delta\theta))$.

We need only leading order in the perturbations, so ABDC is a parallelogram, with area
$r(t)(u+v_-\cos\theta)\delta\theta\delta t$.
We integrate over $\psi$ to get
\begin{equation}
p(v_+,v_-)=-\frac{2r}{\cal A}(v_-\cos\theta+u)\left(\frac{\partial\theta}{\partial v_+}
\right)_{v_-}
\end{equation}
where the final derivative comes because we used $\theta$
to denote the collision variable rather than $v_+$; they are related by Eq.~(\ref{e:v+}).
The factor of two comes from considering both directions for each angle (see above),
and the minus sign from the sign of the partial derivative.  Substituting for $\theta$,
we find
\begin{equation}
p(v_+,v_-)=
\frac{rv_+(v_+^2-v_-^2)}{{\cal A}u\sqrt{8u^2(v_+^2+v_-^2)-(v_+^2-v_-^2)^2-16u^4}}
\end{equation}
Anticipating the expansion in powers of $u$, we now write $v_+=v_-+2su$ so that
$s$ ranges in the fixed interval $[0,1]$, and use this in
the master equation
\begin{eqnarray}
f_t(v,t)&=&\frac{r}{\cal A}\int_0^1\left[p(v,v-2su)f(v-2su,t)\right.\nonumber\\
&&\left.-p(v+2su,v)f(v,t)\right]2uds
\end{eqnarray}
For very small velocities $(v<2|u|)$ we should modify the limits of integration to ensure that
the arguments of $p$ are both positive, however in practice this is not important as we are
interested in long times after which the distribution is almost all at large velocities.

We now consider times of order unity, that is, the period of the oscillations.  The master
equation as it stands is not tractable, being explicitly time-dependent.  Noting again that for
typical particle velocities $v\gg |u|$, we expand
the right hand side of the master equation in a power series in $u$, a
Kramers-Moyal expansion~\cite{RV} as used in Ref.~\cite{DK}.
The functions $f$ and $p$ are expanded
in powers of $u$, which then allows the integral to be performed, leading to
\begin{equation}\label{e:PDE}
f_t=\frac{r}{\cal A}\left[-\pi u(f+vf_v)+\frac{8u^2vf_{vv}}{3}+O(u^3)
\right]
\end{equation}
This is now used to determine $f(v,t)$ at long times.
We note that when $r<r_I$, ie in the IH, IF and FH cases, $\dot{\cal A}/2=-\pi ru$ is the term
that appears in front of the first two terms on the right hand side.  Presumably this term
is also $\dot{\cal A}/2$ for $r>r_I$, as in Ref.~\cite{J}, which gives a comparable
equation; a detailed comparison is given in the Appendix.
Note that terms involving $u^3$ and higher are significant only
for velocities of order $\sqrt{t}$ or less, thus they do not contribute to the main
scaling, which is order $t$.

We now come to the main issue with the oscillating volume.  During
each period, the particles make $O(v)$ collisions with the scatterer during
each of the expanding and contracting phases, thus increasing and decreasing
their speeds by $O(v)$ (with standard deviation $O(\sqrt{v})$).
If we average Eq.~(\protect\ref{e:PDE}) by
neglecting the $u$ terms (which are full time derivatives if the time dependence
of $f+vf_v$ can be ignored) we find $\protect\bar{f}_t=8Cv\protect\bar{f}_{vv}/3$
where $C$ is the average of $u^2/{\cal A}$, leading to
$\protect\bar{f}(v,t)=9ve^{-3v/(8Ct)}/(64C^2t^2)$.  However this cannot capture
the oscillations in $f$.  Thus we propose a more
general ansatz, allowing $v$ to scale with a bounded, but otherwise arbitrary
$2\pi$-periodic function $a(t)$:
\begin{equation}\label{e:ansatz}
f(v,t)=a(t)^2\frac{v}{t}e^{-a(t)v/t}
\end{equation}
where the prefactor $a(t)^2$ is required for normalisation, Eq.~(\ref{e:norm}).
Substituting into Eq.~(\ref{e:PDE})
gives an ODE involving the oscillatory $r$ and $u$:
\begin{equation}
\frac{da}{dt}=\left(t^{-1}-\frac{\pi ur}{\cal A}\right)a-\frac{8}{3t}\frac{u^2r}{\cal A}a^2
\end{equation}
This is a Bernoulli equation, with solution
\begin{equation}
a(t)=t\frac{\sqrt{{\cal A}(t)}}{i(t)},\qquad
i(t)=\frac{8}{3}\int^t \frac{u^2 r}{\sqrt{\cal A}}dt'
\end{equation}
Long time behaviour is characterized by
\begin{equation}
I\equiv \lim_{t\to\infty}\frac{i(t)}{t}=\frac{4}{3\pi}\int_0^{2\pi}\frac{u^2r}{\sqrt{\cal A}}dt,
\end{equation}
an elliptic integral depending on $R$ and $A$ that is easy to evaluate
numerically~\cite{W}.
Thus our final expression for the velocity distribution function is
\begin{equation}\label{e:final}
F(V,t)=\frac{V}{I^2t^2}e^{-V/(It)}
\end{equation}
where $V=v\sqrt{\cal A}$ and $F(V,t)dV=f(v,t)dv$.  In contrast to this exponential
form observed in similar systems~\cite{BCV,DK} the relevant quantity $V$ oscillates rapidly
with respect to the velocities. More generally, in an adiabatic compression we expect the
entropy to vary only slowly.  Indeed for a two dimensional ideal gas with only translational
degrees of freedom entropy per particle is given by~\cite{K46} $R\ln(mTa)+C$ where $R$
and $C$ are constants, $m$ is the mass, $T$ is the temperature and $a$ the area per particle.
Identifying $T$ as proportional to $v^2$ we find that entropy is just the logarithm of $V$ with
some constants.  The importance of the entropy in forced systems was noted in Refs.~\cite{O79};
see also Ref.~\cite{BJS96}.

\begin{figure}
\vspace{0pt}
\centerline{\includegraphics[angle=270,width=230pt]{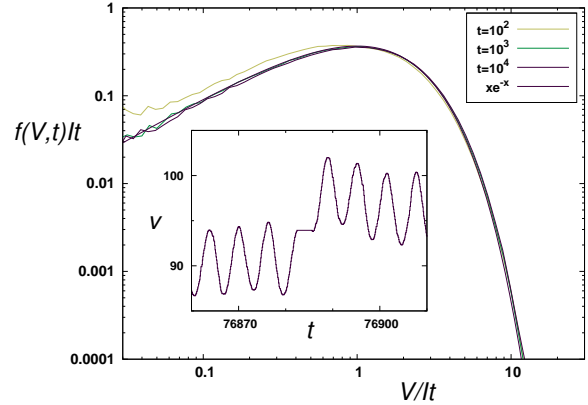}}
\vspace{0pt}
\caption{\label{f:scaling} Convergence to the distribution, Eq.~(\protect\ref{e:final}) for
$R=0.5$ (finite horizon).  Inset: Time dependence of the velocity for a single
trajectory for $R=0.3$ (infinite horizon).  This step function is approximately sinusoidal
except for jumps due to occasional long flights.}
\end{figure}

This distribution is confirmed numerically in Fig.~\ref{f:scaling} for $R=0.5$ and $A=0.03$,
so the FH regime.  For this case numerical integration gives $I\approx0.001327658$.
We simulate the full (not  SWA) time-periodic Lorentz gas. Determining the
time until the next collision thus involves the solution of a transcendental equation, using the
robust quadratically convergent method proposed in~\cite{DL}.
A million initial conditions of particles
are chosen using a Maxwell-Boltzmann distribution at a temperature of $10^{-4}$
consistent with $|u|<A=0.03$.  Particles which are very slow may not collide during the
simulation time, and so delay convergence to the limiting distribution.  Note that
$\cal A$ leads to significant variation in $v$ every cycle; this is clear in the inset, and would be
visible in the distributions for different $t$ if not incorporated correctly.

Next, we consider infinite horizon.  Here, there is no separation between the collision
and oscillation timescales, and the master equation cannot be applied; in general
the velocity distribution is non-universal.  We will however determine
the scaling of velocity with time.  There are two cases, I
(pure infinite horizon) and IF (infinite-finite, also called dynamically
infinite).  For IF the time of free flights is bounded by the
period, but since the velocity can be arbitrarily large, the distance
is unbounded.

For both I and IF we will argue that the FA is of the
form $v\sim t\ln t$.  Physically, more rapid acceleration
is due to the particles making long flights while the
scatterers are contracting, and so missing the cooling phase of the
cycle; see the inset of Fig.~\ref{f:scaling}.
Some will do the opposite and miss the heating phase, but as
with a random walk, the overall effect of larger steps in energy per
cycle is a higher rate of growth of the average energy.

In detail: The probability of a long flight of duration
between $\tau$ and $\tau+d\tau$ is (neglecting multiplicative constants)
of order $(v\tau)^{-3}d(v\tau)=v^{-2}\tau^{-3}d\tau$ for
$\tau\gg v^{-1}$, the typical flight time.  See Ref.~\cite{D}
for an exact constant in the static case.

Collisions normally occur with a rate $\approx v$, so a long flight
avoids a change of velocity $\approx v\tau$ to the particle, since
$u$ is at most of order unity.  Thus the perturbation
$\delta \ln v=\delta v/v$ is of order $\tau$.
The effects are however effectively uncorrelated, so we add variances
in proportion to their probability
\begin{equation}\label{e:logint}
\sum (\delta \ln v)^2\approx \int_{v^{-1}}^1 v^{-2}\tau^{-3} \tau^2 d\tau
\end{equation}
The lower limit of integration is the typical time $v^{-1}$ and the
upper can be taken as the largest time found in the trajectory,
however there is no extra velocity perturbation
for free paths of time greater than the period (of order 1).
Thus we find that per collision the variance scales as
$\ln v/v^2$.  The number of collisions required to reach paths
of order unity is about $v^2/\ln v$, which is less than the
simulation time, noting that in the finite horizon case
FA is of order $v\sim t$.

Each particle thus undergoes a random walk in $\ln v$, taking
a number of collisions $v^2/\ln v$, hence a time $v/\ln v$ to
take each step.  The total time for the trajectory is dominated
by the largest value of $v$ in the path, so that the velocity
is typically of order $t\ln t$.  This argument follows through
for both infinite (I) and infinite/finite (IF) cases, although
it is more pronounced (larger coefficient) in the former.  Note
the paradoxical effect in which the intermittency leads to
long times without collisions, but is responsible for
increasing the (collision-driven) FA.

We remark that the transport in velocity in both finite
and infinite horizon cases is purely diffusive, ie free of a drift
term, which would markedly alter the exponent of $t$, in
contrast to the numerical results presented here. This means the
dynamics is almost certainly recurrent, eventually returning near
its starting point on very long time scales.

\begin{figure}\vspace{0pt}
\centerline{\includegraphics[angle=270,width=230pt]{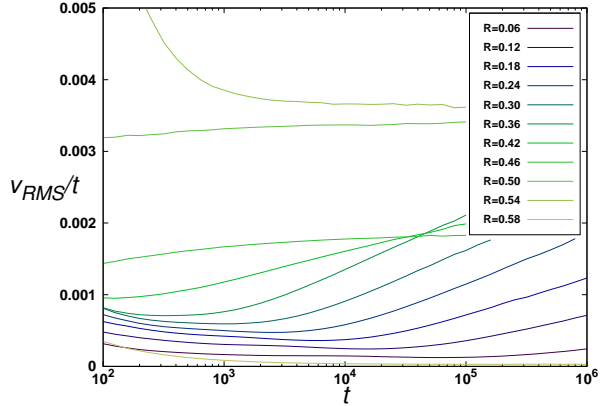}}
\vspace{0pt}
\caption{\label{f:mom}  Velocity growth for different $R$.  The curves start horizonal ($v\sim t$),
and those with $R<0.45$, where the infinite horizon is noticeable, then increase linearly
($v\sim t\ln t$).  The lowest curve, $R=0.58$, is a billiard completely confined for all time,
exhibiting the weakest Fermi acceleration.}
\end{figure}

The dependence of the FA on the radius (and hence the
finite/infinite horizon status) is shown in Fig.~\ref{f:mom}.
The time regimes are (a) dominated by the initial Maxwell-Boltzmann
distribution, (b) linear growth of $v$, (c) for I, increase as the
logarithmic Eq.~(\ref{e:logint}) starts to dominate the normal
linear acceleration when there start to be several collisions
per oscillation cycle.  The linear growth in $v$ is proportional to the
cross-section (roughly $R$), thus the collision rate is roughly $R^2t$
and the transition to several collisions per cycle occurs at roughly
$t\sim R^{-2}$ for small $R$,
as can be seen from the minima in Fig.~\ref{f:mom}. 

To summarise, we have demonstrated several new methods and effects for
systems with periodic volume oscillations.  The master equation
approach can be applied to any time-dependent container for which the
static dynamics is chaotic, specifically with integrable decay of correlations.
The intermittent case, for example with non-integrable decay of correlations,
does not appear to exhibit a universal velocity distribution and so needs
further study.  

The FC parameter range, in which the particles are alternately
confined and unconfined, is also unexplored.  In particular, it
would be interesting to investigate the many rapid collisions undertaken
by a particle near where the two scatterers touch, a new
``dynamical cusp'' mode of intermittency.  The
possibility of an unbounded number of collisions suggests that
since in each (approximately perpendicular) collision, a fixed
quantity $u$ is added to the velocity, the velocity itself can
become unbounded in finite time for a small set of initial
conditions, a further example of intermittency enhanced acceleration.

Finally, physical experiments involve many other features - soft potentials,
external fields, interparticle interactions and quantum effects.  Our results suggest
a thermodynamic approach, characterising particle distributions in terms of entropy.

\paragraph{Acknowledgments}
The authors are grateful for support from Brazilian agencies
FAPESP, FUNDUNESP and CNPq.  This research was supported by
resources supplied by the Center for Scientific Computing
(NCC/GridUNESP) of the S\~ao Paulo State University (UNESP).

\paragraph{Appendix}
Here we compare Eq.~(\protect\ref{e:PDE}) with the previous result of Ref.~\cite{J}.
The $f_{vv}$ term in Eq.~(\protect\ref{e:PDE}) is of the same form as the diffusive term in
Ref.~\protect\cite{J} (substituting energy $E=mv^2/2$ and mass $m=1$), but
has a different coefficient, for two reasons:  Ref.~\protect\cite{J} neglects the effect of
motion of the boundary on the collision rate (``aberration,'' Fig.~\ref{f:diag}).  In addition
we assume independence of collisions, good for the Lorentz gas except very close to $r_I$.  
In detail, Eqs.~(3.12,~3.16a,~3.22a) of Ref.~\protect\cite{J} give the same as in
Eq.~(\protect\ref{e:PDE}) but with $8u^2/3$ replaced by $4\sum_{j=-\infty}^{\infty}c_j$ where
$c_j$ is the autocorrelation of the function $u\cos\theta-\protect\langle u\cos\theta\protect\rangle$
in our notation, and here $u$ is independent of position.
The main term is $4c_0=(8/3-\pi^2/4)u^2$, given (correctly) in Eq.~(A7) of Ref.~\cite{J}.
The other correlations are small, for example, the largest term for $r=0.53$, just less than
$r_I$, is $4c_3\approx 0.008$. Numerical simulations (Fig.~\protect\ref{f:scaling})
are consistent with $8/3$ plus
undetectable correlation corrections, but not with $8/3-\pi^2/4$.  If desired, we can incorporate
the other $c_j$ into our approach directly, by increasing the coefficient to $8/3+4\sum_{j\neq 0}c_j$.

\bibliographystyle{custom}
\bibliography{tdep}

\end{document}